\documentclass[12pt,preprint]{aastex}
\usepackage[dvips]{color}
\usepackage{graphicx}

\def\msol{\hbox{\kern 0.20em $M_\odot$}}
\def\lsol{\hbox{\kern 0.20em $L_\odot$}}
\def\rsol{\hbox{\kern 0.20em $R_\odot$}}
\def\sr{\hbox{\kern 0.20em sr}}
\def\srmu{\hbox{\kern 0.20em sr$^{-1}$}}

\def\g{\hbox{\kern 0.20em g}}
\def\gmu{\hbox{\kern 0.20em g$^{-1}$}}
\def\kg{\hbox{\kern 0.20em kg}}
\def\pc{\hbox{\kern 0.20em pc}}

\def\mum{\hbox{\kern 0.20em $\mu$m}}
\def\mumd{\hbox{\kern 0.20em $\mu$m$^{-2}$}}
\def\cm{\hbox{\kern 0.20em cm}}
\def\m{\hbox{\kern 0.20em m}}
\def\km{\hbox{\kern 0.20em km}}
\def\nm{\hbox{\kern 0.20em nm}}

\def\s{\hbox{\kern 0.20em s}}
\def\h{\hbox{\kern 0.20em h}}
\def\sec{\hbox{\kern 0.20em sec}}
\def\min{\hbox {\kern 0.20em min}}
\def\smu{\hbox{\kern 0.20em s$^{-1}$}}
\def\smd{\hbox{\kern 0.20em s$^{-2}$}}
\def\an{\hbox{\kern 0.20em an}}
\def\anmu{\hbox{\kern 0.20em an$^{-1}$}}
\def\deg{\hbox{\kern 0.20em $^{\rm o}$}}
\def\yr{\hbox{\kern 0.20em yr}}
\def\yrmu{\hbox{\kern 0.20em yr$^{-1}$}}
\def\Myr{\hbox{\kern 0.20em Myr}}
\def\Mymu{\hbox{\kern 0.20em Myr$^{-1}$}}
\def\K{\hbox{\kern 0.20em K}}
\def\pcmu{\hbox{\kern 0.20em pc$^{-1}$}}
\def\pcmd{\hbox{\kern 0.20em pc$^{-2}$}}
\def\pcmt{\hbox{\kern 0.20em pc$^{-3}$}}
\def\kms{\hbox{\kern 0.20em km\kern 0.20em s$^{-1}$}}
\def\kmpd{\hbox{\kern 0.20em km$^{2}$}}
\def\kpc{\hbox{\kern 0.20em kpc}}
\def\cms{\hbox{\kern 0.20em cm\kern 0.20em s$^{-1}$}}
\def\erg{\hbox{\kern 0.20em erg}}
\def\ergs{\hbox{\kern 0.20em erg}}
\def\cmpd{\hbox{\kern 0.20em cm$^2$}}
\def\cmmd{\hbox{\kern 0.20em cm$^{-2}$}}
\def\cmms{\hbox{\kern 0.20em cm$^{-6}$}}
\def\cmpt{\hbox{\kern 0.20em cm$^3$}}
\def\cmmt{\hbox{\kern 0.20em cm$^{-3}$}}
\def\mpd{\hbox{\kern 0.20em m$^2$}}
\def\mmd{\hbox{\kern 0.20em m$^{-2}$}}
\def\mpt{\hbox{\kern 0.20em m$^3$}}
\def\mmt{\hbox{\kern 0.20em m$^{-3}$}}
\def\mujy{\hbox{\kern 0.20em $\mu$Jy}}
\def\mjy{\hbox{\kern 0.20em mJy}}
\def\Mj{\hbox{\kern 0.20em MJy}}
\def\jy{\hbox{\kern 0.20em Jy}}
\def\ghz{\hbox{\kern 0.20em GHz}}
\def\srmd{\hbox{\kern 0.20em sr$^{-1}$}}
\def \kms{km~$\rm{s}^{-1}$}

\def \mum{$\mu$m}

\def\G{\hbox{\kern 0.20em G}}

\def\h13cop{\hbox{H$^{13}$CO$^{+}$}}

\def\S+{\hbox{S{\small II}}}

\slugcomment{The Evolving ISM in the Milky Way \& Nearby Galaxies}

\shorttitle{The IMF} \shortauthors{YourName et al.}

\begin{document}

\newcommand{\jfourteen}{\hbox{$J=14\rightarrow 13$}}
 \title{The Stellar Initial Mass Function in 2007: A Year for Discovering Variations}

\author{Bruce G. Elmegreen\altaffilmark{1},
}

\altaffiltext{1}{IBM T.J. Watson Research Center, Yorktown Hts., NY
10598; bge@us.ibm.com}

\begin{abstract}
The characteristic mass ($M_c$) and slope ($\Gamma$) of the IMF are
reviewed for clusters, field regions, galaxies, and regions formed
during cosmological times. Local star formation has a somewhat
uniform $M_c$ and $\Gamma$. Statistical variations in $\Gamma$ are
summarized, as are the limitations imposed by these variations.
Cosmological star formation appears to have both a higher $M_c$ and
a slightly shallower slope at intermediate to high stellar mass. The
center of the Milky Way may have a shallow slope too. Field regions
have slightly steeper slopes than clusters, but this could be the
result of enhanced drift of low mass stars out of clusters and
associations. Dwarf galaxies also have steeper slopes. Results from
the observation of pre-stellar clumps are reviewed too. Pre-stellar
clumps appear to have about the same mass function as stars and are
therefore thought to be the main precursors to stars. If this is the
case, then the IMF is generally determined by gas-phase processes.
Brown dwarf formation also shares many characteristics of star
formation, suggesting that they form by similar mechanisms.
\end{abstract}


Posted on the conference web page,
http://ssc.spitzer.caltech.edu/mtgs/ismevol/, for the 4th Spitzer
Science Center Conference, ``The Evolving ISM in the Milky Way and
Nearby Galaxies: Recycling in the Nearby Universe;'' December 2-5, 2007
at the Hilton Hotel, Pasadena, California.

\lefthead{Bruce G. Elmegreen}

\righthead{The Initial Stellar Mass Function}

\section{Introduction}

The year 2007 has been a watershed for the discovery IMF variations.
This review includes observations of the characteristic mass, IMF
slope, field star IMF, and pre-stellar clump mass functions, along
with variations in these quantities that have come to light
recently. Theoretical considerations are discussed briefly.

\section{A Characteristic Mass}

\begin{figure}[h]
\centerline{
\includegraphics[width=230pt]{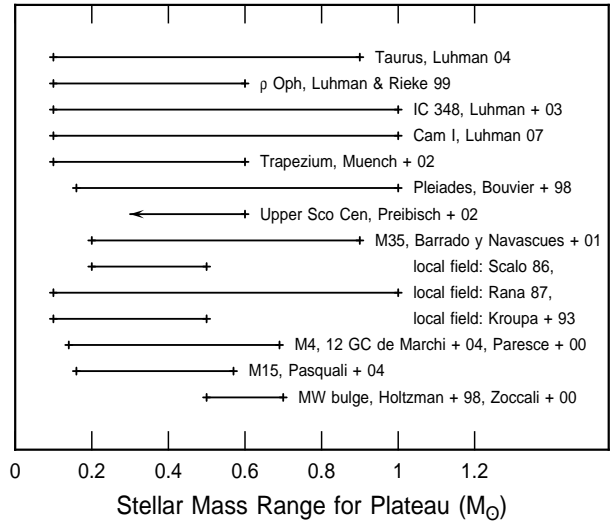}
} \caption{\label{imf_sst_peak} Mass ranges for the plateau in the
IMF for several clusters and regions of star formation. The plateau
is the mass range for the peak of the IMF plotted on a log-log
scale, below the power law part characterizing intermediate to high
mass stars, and above the apparent drop off in the brown dwarf
range.}
\end{figure}

The IMF is usually fitted to a power law at high mass ($M>1\;
M_\odot$) and a rollover at low mass. The peak in the IMF on a
log-log plot is at $\sim0.3\; M_\odot$. We think of this as a
characteristic mass for star formation, $M_c$. Globular clusters and
the Milky Way bulge have about the same mass range in the IMF
plateau as local clusters. Figure \ref{imf_sst_peak} summarizes a
sample of observations. A wide variety of clusters with diverse
environments, locations, metallicities and ages have the same $M_c$.
For example, cluster Blanco 1 is like the Pleiades in age and mass
(Moraux et al. 2007), but Blanco 1 is less dense, lies 240 pc off
the plane, and has subsolar abundances for [Ni/Fe],[Si/Fe],
[Mg/Fe],[Ca/Fe]; still, the IMFs of these two clusters are almost
identical. Digel 2N and S are two far-outer Galaxy clusters
($R_{gal}\sim19$ kpc), yet they have normal IMFs, like Orion's, for
an age of 0.5-1Myr (Yasui et al. 2007).

The same characteristic mass occurs for low density and low pressure
regions like Taurus, and high density and high pressure regions like
globular clusters (which formed as super star clusters). This
implies that the conditions which determine the characteristic mass
vary together.  If $M_c\sim M_J$ for thermal Jeans mass $M_J$ (Clark
\& Bonnell 2005; Bate \& Bonnell 2005), then $M_J\sim
T^{3/2}G^{-3/2}\rho^{-1/2}$ is constant and so $T\propto\rho^{1/3}$
for temperature $T$ and density $\rho$.  If $M_J$ is determined at
the point of grain-gas thermal coupling during collapse (Whitworth
et al 1998; Larson 2005; Jappsen et al. 2005), which corresponds to
some high density, $\rho_{coupling}$, then T has to scale with
$\rho_{coupling}^{1/3}$. Heating from outside and inside a
star-forming cloud must balance cooling so that $M_J\sim$constant.
How is this possible? Perhaps $M_J$ is not involved after all. $M_c$
could also depend on core sub-fragmentation and protostellar
feedback.  A recent discussion of what could conspire
to make $M_J$ constant is given in Elmegreen, Klessen \& Wilson
(2008).

There are recent reports of observations that find different values
for $M_c$:
\begin{itemize}
\item Getman et al. (2007) studied triggered star formation in the
cometary globule IC 1396N and suggested that the average x-ray
luminosity is high for the small number of stars present (e.g.,
compared to Ophiuchus and Serpens). They concluded that the region
could have had preferential triggering of massive stars, which means
high $M_c$.
\item On a galactic scale, van Dokkum (2008) studied the
mass-to-light ratio and U-V for early type galaxies at redshifts in
the range $0.02<z<0.83$. He noted that the time dependent variation
of M/L depends on the IMF slope near $M=1\;M_\odot$, which is the
mass range that currently dominates the flux. The best fit model had
an approximately flat IMF at $1\;M_\odot$, which means the plateau
has to include this mass. He also noted that this result is
consistent with Balmer absorption strengths in early type galaxies
as a function of redshift.
\item Dav\'e (2008) found that the star
formation rate per unit mass measured for numerous galaxies in a
wide mass range ($10^{9.4}-10^{12}\;M_\odot$) agrees with that from
simulations at $z\sim0$ but exceeds the simulation predictions for
$z\sim2$. This was explained as a result of increasing IMF
characteristic mass with $z$. High $M_c$ produces a population with
a lot of luminosity from star formation but not much build up of
mass. He considered the ratio of the high-mass star formation rate
to the stellar mass for an evolving IMF compared to a standard IMF
and suggested that $M_c=0.5\left(1+z\right)^2$, i.e., the
characteristic mass increases monotonically with redshift and was
$\sim9$ times higher at $z=2$.
\item Fardal et al. (2007) considered constraints from cosmic
background starlight and the local luminosity density of galaxies.
All of the commonly-observed local IMFs failed to reproduce the
observations, while a ``Paunchy'' IMF, shifted toward higher mass,
worked well.
\item Komiya et al. (2007) found a high characteristic
mass in the IMF of extreme metal-poor (EMP) Milky Way stars. The IMF
is constrained by the fractions of EMP stars that are C-rich with
s-process and those that are C-rich without s-process. C-rich EMP
stars are probably surviving low-mass binary members (models of
stellar evolution and binary mass transfer are involved). The
theoretical ratio of C-rich-no-s-process to C-rich s-process stars
agrees with the observations only if $M_c\sim5\;M_\odot$.  The
theoretical fraction of EMP stars that are C-rich s-process also
agrees if $M_c\sim5\;M_\odot$. These metal poor stars have
$[Fe/H]<-2.5$.
\end{itemize}

To summarize: local IMFs (mostly in clusters) have a nearly
universal $M_c$. This requires some conspiracy of conditions if
$M_J$ is involved.  Cosmological evidence suggests $M_c$ was higher
in the past, by a factor of $\sim10$, shifting from
$\sim0.3\;M_\odot$ today to $\sim5\;M_\odot$ at $z\sim2$ or more.
These are not Population III stars, but normal stars that show up in
galaxies today, including the Milky Way. This conclusion about
increasing $M_c$ comes from considerations of the mass-to-light
ratio and U-V for ellipticals, from the star formation rate per unit
mass versus redshift, from cosmic background starlight and the local
luminosity density, and from extreme metal-poor stars in the Milky
Way.

\section{IMF Slope}
\subsection{Observations}

Scalo (1998) summarized various IMF slopes and plotted them as a
function of the average mass observed. There was a large scatter in
the slope for each mass range, but the general trend and average was
consistent with a Salpeter IMF (slope $\Gamma=-1.35$ on a log-log
plot) at high mass and a flattening of the IMF (slope$\sim0$) around
and below $0.3\;M_\odot$. Many dense clusters have Salpeter IMFs:
R136 in the 30 Dor region of the LMC (Massey \& Hunter 1998), h and
$\chi$ Persei (Slesnick, Hillenbrand \& Massey 2002), NGC 604 in M33
(Gonzalez Delgado \& Perez 2000), NGC 1960 and NGC 2194 (Sanner et
al. 2000), NGC 6611 (Belikov et al. 2004), to name a few. To
consider some extreme cases, Westerlund 2, with $\sim5000$ stars
observed, containing $\sim7000\; M_\odot$, 2 Myr old, at 2.8 kpc
distance and behind 5.8 mag extinction, and with mass segregation,
has an IMF slope on average that is $\Gamma=-1.2\pm0.16$, i.e.,
approximately Salpeter (Ascenso et al. 2007). NGC 346 in the SMC,
which has $\sim1/5$ solar metallicity, has a Salpeter IMF on average
with a radial gradient in the slope (Sabbi et al. 2007). The IMF of
the Rosette cluster derived from x-ray observations looks like the
Orion IMF from its x-ray stars (Wang et al. 2007).

\begin{figure}[h]
\centerline{
\includegraphics[width=200pt]{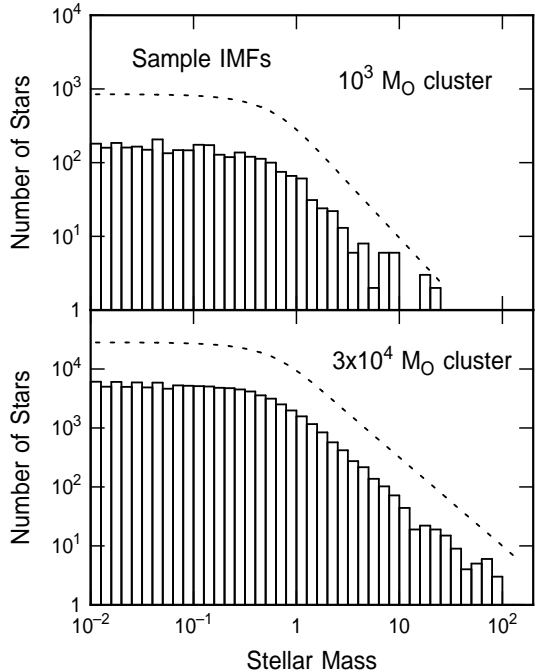}
} \caption{\label{imf3_sst} Two randomly generated IMFs based on
sampling the analytical IMF given by equation 1 with $\Gamma=1.5$.
The top IMF has fewer stars than the bottom IMF, as shown in the
label.  The break up of the smooth histogram at high mass is to the
same degree in each panel because the number of stars close to the
maximum likely stellar mass is the same.}
\end{figure}

%[width=200pt,height=200pt]

However, the IMF in the giant Milky Way cluster NGC 3603 is
relatively flat with a slope of $\Gamma=-0.74$ (Harayama et al.
2007). There are $\sim7500$ stars measured, so the uncertainty in
the slope from counting statistics is only $\pm0.02$. Mass
segregation from dynamical evolution is not significant as there is
no IMF steepening beyond the half-mass radius. The uncertainty in
the slope from binaries is $\pm0.04$. A similar result was found by
Stolte et al. (2006), who got a slope of $\Gamma=-0.9\pm0.15$,
shallower than the Salpeter slope of $-1.35$. On the other hand,
Preibisch et al. (2002) found a steep IMF in the Upper Sco-Cen
star-forming region, which is not a bound cluster. They derived
slopes of $\Gamma=-1.8$ for 0.6 to $2\;M_\odot$ stars and $-1.6$ for
2 to $20\;M_\odot$ stars.  W51 has a spatially varying IMF with a
mean slope of $-1.8$ in four subgroups but statistically significant
excesses in the numbers of the most massive stars in 2 subgroups
(Okumura et al. 2000).

\begin{figure}[h]
\centerline{
\includegraphics[width=250pt]{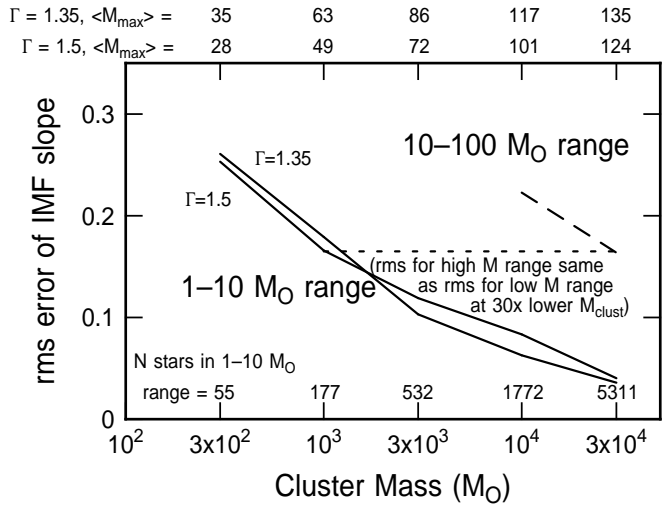}
} \caption{\label{imf3_sst_sum} The rms deviation in the slope of
the IMF is shown versus the mass of the cluster (bottom axis) and
the maximum likely mass of the stars (top axis). The two solid lines
are for two values of $\Gamma$, the intrinsic slope (Eq. 1). The rms
is evaluated in two mass ranges, from 1 to $10\;M_\odot$ and from 10
to $100\;M_\odot$ (for the high mass range, $\Gamma=1.5$). Only high
mass clusters have a well-sampled IMF in the upper stellar mass
range. The rms in the high-mass slope of a $3\times10^4\;M_\odot$
cluster is the same as the rms in the intermediate-mass slope of a
$10^3\;M_\odot$ cluster. The number of stars in the $1-10\;M_\odot$
mass range is indicated along the bottom axis for the $\Gamma=1.5$
case.}
\end{figure}

\begin{figure}[h]
\centerline{
\includegraphics[width=350pt]{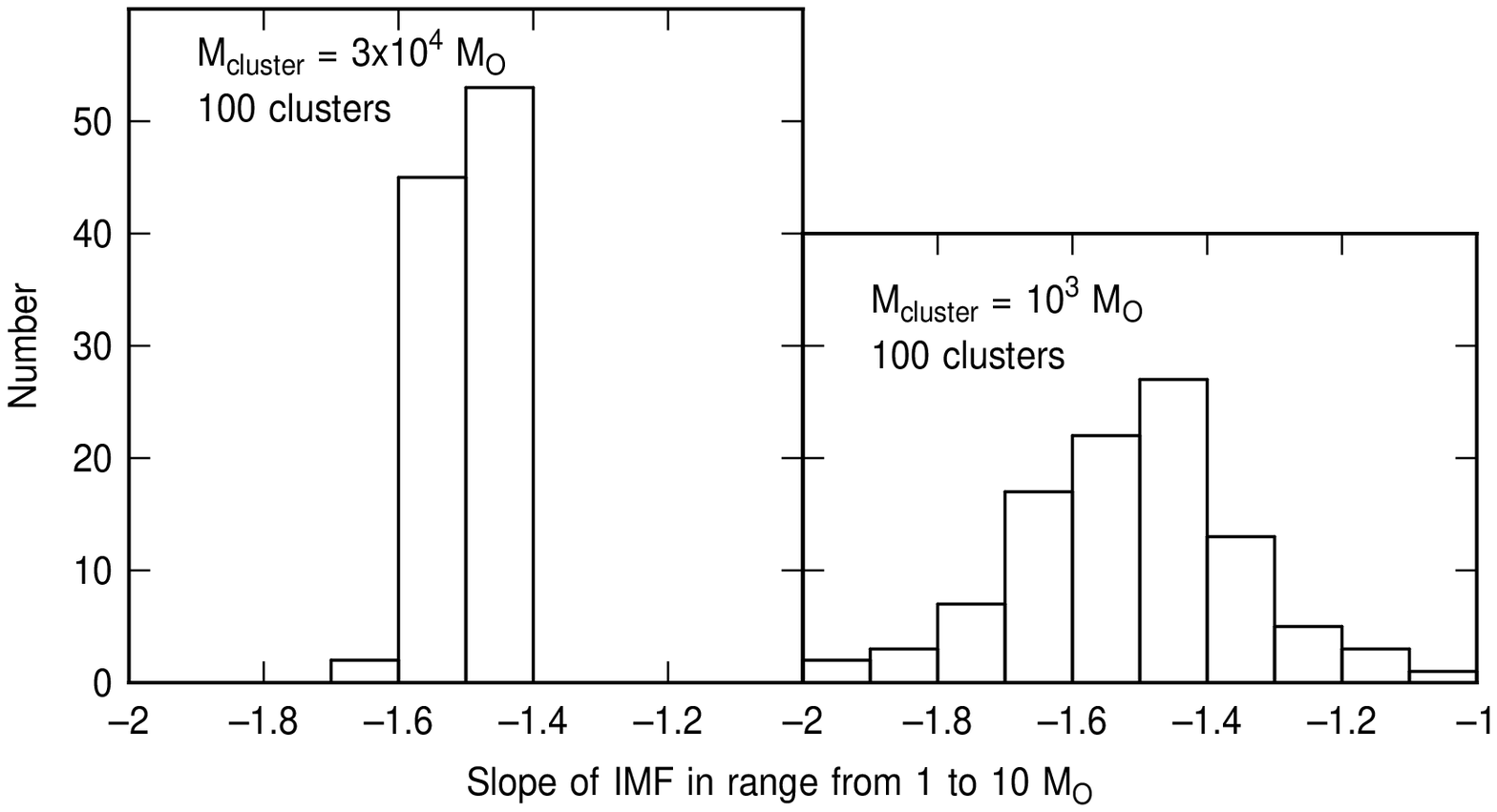}
} \caption{\label{imf_sst_his} Histograms of IMF slopes for
$3\times10^4\;M_\odot$ and $10^3\;M_\odot$ clusters. An IMF with
$\Gamma=-1.5$ was used, which reflects the mean value in these
histograms. The rms plotted in the previous figure is the breadth of
the histogram here for the same cluster mass. }
\end{figure}
\subsection{What should the statistical rms in the slope be?}

The expected rms can be found by randomly sampling the IMF in a
model cluster. We assume model IMFs, $n(M)$, in equal intervals of
mass:
\begin{equation}
n(M)dM=M^{-(\Gamma+1)}(1-\exp^{-(M/M_t)^{\Gamma}})dM\end{equation}
where $\Gamma=1.5$ and 1.35 (Salpeter function), which is a power
law having a turnover at $M_t=0.5\;M_\odot$. We assume a maximum
stellar mass $M_{max}=150\;M_\odot$.

Figure \ref{imf3_sst} shows two randomly generated IMFs, one for a
cluster with $10^3\;M_\odot$ and another for a cluster with
$3\times10^4\;M_\odot$. Clearly the larger cluster has a smoother
IMF at low mass because statistical variations are smaller. If we
sample the IMFs for 100 clusters of the same mass, we can find the
slope for each cluster in a certain mass range, and then find the
rms deviations of these slopes around the average value. Figure
\ref{imf3_sst_sum} shows the rms deviations of the slopes versus the
cluster mass. The pair of lines ($\Gamma=1.35$ and 1.5) is for the
stellar mass interval between 1 and $10\;M_\odot$ and the single
dashed line ($\Gamma=1.5$) is for the $10-100\;M_\odot$ range. The
maximum likely stellar mass in the cluster is shown on the top axis
for each assumed IMF slope $\Gamma$.

For $10^3\;M_\odot$ clusters, corresponding to a maximum stellar
mass of $\sim50\;M_\odot$, the $1-\sigma$ uncertainty in the IMF
slope in the $1-10\;M_\odot$ range is $\sim0.17$. In the
$10-100\;M_\odot$ range, the IMF uncertainty drops below 0.2 only
for clusters more massive than $10^4\;M_\odot$, corresponding to a
maximum stellar mass of $>100\;M_\odot$.  The rms for the high
stellar mass range in a massive cluster is the same as the rms for
the low stellar mass range in a cluster that has 30 times lower
mass.

\begin{figure}[h]
\centerline{
\includegraphics[width=200pt]{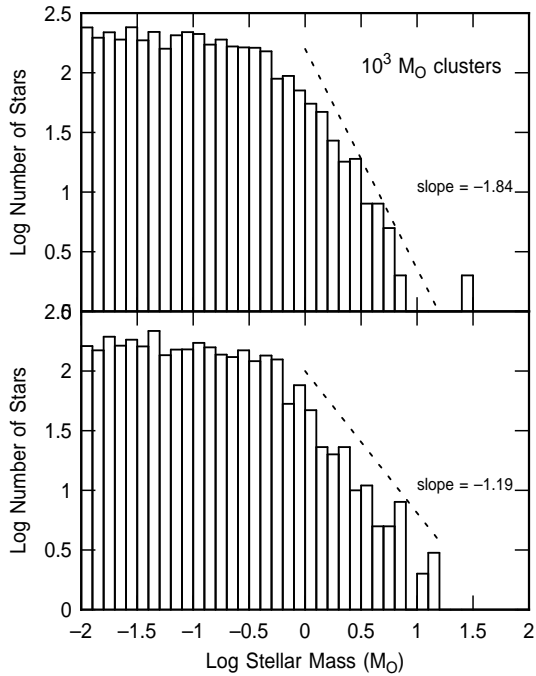}
} \caption{\label{imf_sst_imf} Two IMFs in extreme cases of the same
randomly sampled model.}
\end{figure}

Figure \ref{imf_sst_his} shows histograms of the IMF slopes for two
cluster masses. For a $10^3\;M_\odot$ cluster, which is a typically
massive cluster in the Milky Way, like NGC 3603, the IMF slope
varies from $\Gamma=-1.8$ to $-1.1$. Sample IMFs are shown for these
two extremes in Figure \ref{imf_sst_imf}. They look like reputable
IMFs without a large amount of scatter from mass interval to mass
interval, but their slopes are clearly different. This difference is
entirely random. This example illustrates how the {\it intrinsic}
IMF slope cannot be determined to within several tenths for even the
most massive clusters in the Milky Way.

The implications of this exercise are worth noting. The upper mass
decade for every cluster has an IMF slope uncertainty of $\sim0.2$
or more because there is always the same small number of stars in
this upper decade ($<500$). When the cluster mass is so large
($\sim10^5\;M_\odot$) that the maximum possible stellar mass is
reached, the number of stars in the upper mass decade goes up and
the uncertainty goes down. Second, curvature in the high mass part
of the IMF will be difficult to observe or rule out for individual
clusters. Similarly, real upper mass IMF variations from cluster to
cluster are undetectable.

No cluster IMF has ever been observed throughout the whole stellar
mass range. The upper mass IMF needs a very massive cluster, like 30
Dor, massive clusters are rare, and the nearest is too far away to
see the low mass stars. On the other hand, the nearest clusters,
which are required to study brown dwarfs and low mass stars, are the
most common clusters and they are all low mass clusters, having few
high mass stars. Until we can observe the lowest mass stars in the
highest mass clusters, an "IMF" makes sense only for an ensemble of
clusters or stars.

\section{Is the Field a good place to measure the IMF?}

The field is peculiar because there is no star formation there now.
A decreasing star formation rate steepens the IMF if the decrease is
not accounted for (Elmegreen \& Scalo 2006).  The field also
receives the evaporated stars (low mass stars) from clusters, and
the longest-lived stars (low mass stars) from associations, making
the field IMF steeper than in clusters even if the total galaxy IMF
is the same as in clusters.

Field IMFs can include so many stars that rms errors in the slope
are negligible. For example, the field IMF for the LMC measured by
Parker et al. (1998) contained 37,300 stars and had a mean slope of
$\Gamma=-1.80\pm0.09$ for $M > 2\;M_\odot$.  Massey et al. (1995,
2002) found a steep IMF in the remote fields of the LMC
($\Gamma=-4$) and SMC ($\Gamma=-3.6$), more than 30 pc from a Lucke
\& Hodge (1970) or Hodge (1986) association. Their IMF was complete
down to $25\;M_\odot$ and they assumed a constant star formation
rate over the last 10 My; 450 stars were in the in LMC sample, for
which we would predict an rms uncertainty of $\pm0.15$, which is
relatively small. Another field region near LMC4 has a slope of
$\Gamma=-5$ for $M=0.9-2\;M_\odot$ and $\Gamma=-2.6$ for
$M=0.9-6\;M_\odot$ (Gouliermis et al. 2005). However, a field region
near 30 Dor has the Salpeter IMF: $\Gamma=-1.38\pm0.04$ for
$M=7-40\;M_\odot$ (Selman \& Melnick 2005).

Chandar et al. (2005) compared field IMF spectra to evolution models
in nearby starburst galaxies, often finding a steeper IMF slope than
Salpeter, by $\sim0.5$, or a low maximum stellar mass as an
alternative explanation. Similarly \'Ubeda et al. (2007a,b) found a
steep IMF in the dwarf galaxy NGC 4214, $\Gamma=-1.83\pm0.07$, by
counting main sequence stars in intervals of the HR diagram. Field
and cluster IMFs were not much different in that study. They
suggested that $\Gamma$ could be even steeper than this. Also, from
IUE spectra in this galaxy, Mas-Hesse \& Kunth (1999) suggested
$\Gamma = -2.0$.

Hoversten \& Glazebrook (2007) fit a star formation model to
H$\alpha$ equivalent widths and g-r colors in 140,000 SDSS galaxies,
minimizing over variations in $\Gamma$, metallicity, ages, and star
formation history. They found an IMF slope that got steeper for
lower mass galaxies, ranging from the Salpeter value for $M_R=-22$
mag to $\Gamma=-1.6$ for $M_R=-17$ mag.

The Milky Way and M31 bulge [Fe/H] abundances suggest shallow IMFs
at intermediate to high mass, $\Gamma\sim-1$ to $-1.1$ (Ballero et
al. 2007ab). The central region of the Galaxy may have a shallow
current-day IMF too, $\Gamma=-0.85$ (Paumard et al. 2006), based on
40 OB supergiants, giants, and main-sequence stars in 2 rotating
disks within 0.5 pc of the nucleus; the ages of these stars are
$6\pm2$ Myr and the total mass is $1.5\times10^4\;M_\odot$. Also in
the center, 329 late-type giants within 1 pc of SgrA* have
$\Gamma=-0.85$ for 12 Gyr of SF (Maness et al 2007). In a third
study of the Galactic center, Nayakshin \& Sunyaev (2005) found few
x-ray stars associated with the massive stars orbiting Sgr A* and
proposed in situ star formation with a shallow IMF.  The IMF in the
Arches cluster based on deep AO images is also slightly flatter than
Salpeter, $\Gamma=-1$ to $-1.1$; Fokker-Planck models were used to
account for mass segregation over the age of the cluster (Kim et al.
2006).

What about ``top heavy'' IMFs in super star clusters? There are
several observations that suggest this. Sternberg (1998) found a
high value of L/M in NGC 1705-1 that suggests either $|\Gamma|<1$ or
an inner mass cutoff for stars. Smith \& Gallagher (2001) found the
same for M82F: a high inner cutoff of 2 to $3\;M_\odot$ if
$\Gamma=-1.3$. Alonso-Herrero et al. (2001) found a high L/M in the
starburst NGC 1614.  McCrady et al. (2003) suggested that MGG-11 in
M82 is deficit in low mass stars.  Mengel et al. (2002) got the same
in NGC 4038/9. But other super star clusters have normal IMFs: NGC
1569-A (Ho \& Filippenko 1996; Sternberg 1998), NGC 6946 (Larsen et
al. 2001), M82 MGG-9 (McCrady et al. 2003).  It is unclear if any of
these super star cluster IMFs are correct. Bastian \& Goodwin (2006)
found that all of the odd IMFs in SSCs are in the youngest clusters,
and suggested that these young clusters may not be relaxed enough to
give an accurate dynamical mass.

There are other IMF oddities too. Massive elliptical galaxies have
slightly flatter-than-Salpeter IMFs in studies by Pipino \&
Matteucci (2004) and Nagashima et al. (2005b). Clusters of galaxies
suggest a history of flat IMFs in elliptical galaxy bursts (Renzini
et al. 1993; Loewenstein \& Mushotsky 1996; Chiosi 2000; Moretti,
Portinari, \& Chiosi 2003; Tornatore et al. 2004; Romeo et al. 2004;
Portinari et al. 2004; Nagashima et al. 2005a). Low surface
brightness galaxies may have steep IMFs, $\Gamma\sim-2.9$ (Lee et
al. 2004; Hoversten \& Glazebrook 2007; \'Ubeda et al. 2007). The
red halos in BCD galaxies and stacked halos around disks have
$\Gamma=-3.5$ (Zackrisson, et al. 2004). Also, more locally, there
are apparently bare or single-forming O stars that represent 4\% of
O stars (deWit et al. 2005), and there are O-stars on the periphery
of clusters, possible triggered or not mass segregated.

In summary, the Salpeter slope ($\Gamma=-1.35$) typically occurs in
clusters with marginally insignificant variations. A steeper IMF
seems to be the rule for field regions, although there are
uncertainties in the star formation history and there are possibly
important effects from low-mass stellar drift out of nearby clusters
and associations. Slightly steeper IMFs are found for low mass or
low surface brightness galaxies, $\Gamma\sim-1.5$ to $-2.9$.
Slightly flatter IMFs are found for extremely active star-forming
regions in the Milky Way and M31 bulges, the Milky Way nucleus, and
bursting ellipticals, all of which have $\Gamma\sim-1$.

\section{Theory: Origins of Stars and the IMF}

There are three types of theories for the IMF:
\begin{itemize}
\item Fragmentation (``top down''), most likely driven by
turbulence \& self-gravity (``turbulent fragmentation'').
Simulations of this process have been studied by several teams
(e.g., Tilley \& Pudritz 2007; Padoan et al. 2007; Li et al. 2004;
Nakamura \& Li 2005, 2007; Martel, Evans \& Shapiro 2006; see review
by MacLow \& Klessen 2004). This process is most clearly related to
the initial phases of star formation, and may also be most relevant
to intermediate stellar masses, possibly including the IMF turnover
at $M_c\sim0.3\;M_\odot$ It may be what connects the characteristic
mass, $M_c$ to the thermal Jeans mass $M_J$.

\item Accretion and coagulation ("bottom up"), where clumps accrete from
filaments and sheets to grow into stars (including ``competitive
accretion''; see review by Bonnell et al. 2007). Also included in
these models are situations where clumps or protostars coagulate
inside dense clusters. This may be most relevant to the late stages
in star formation, and to the most massive stars. Coagulation or
capture should form some binaries, and protostar interactions should
influence the disk fraction.

\item ``Interruption'' to fragmentation and accretion, where
either of the previous two processes gets interrupted so the final
stellar mass is smaller than it would have been.  Included here are
models for core ejection from dense groups, and core ionization
(Boss 2001, Reipurth \& Clarke 2001, Bate et al. 2002; Whitworth \&
Zinnecker 2004; Whitworth \& Goodwin 2005; Goodwin \& Whitworth
2007; Umbreit et al. 2005; and others). This process may be most
relevant to brown dwarfs.
\end{itemize}

\section{Brown Dwarfs: Fragmentation or Ejection?}

Several observations suggest that brown dwarfs and stars form by the
same mechanism. First, brown dwarfs and stars have the same spatial
distribution in Taurus (Briceno et al. 2002; Luhman 2004a, 2006),
whereas ejection in the ``interruption'' model should produce more
dispersed brown dwarfs than stars (Kroupa \& Bouvier 2003). Second,
the disk fraction for brown dwarfs is about the same as the disk
fraction for stars, and also about the same as the fraction of
isolated planetary mass objects ($\sim30$\%; Luhman et al. 2006;
Scholz \& Jayawardhana 2007). This is also contrary to the ejection
scenario, which might be expected to produce fewer or smaller disks
in brown dwarfs. Third, there is a binary brown dwarf in Cha I with
a 240 AU separation; this is a fragile system that could probably
not have been ejected (Luhman 2004b). Fourth, the accretion rate
scales with protostellar mass as $M^{2.1}$, spanning the stellar to
brown dwarf boundary without a change (Muzerolle et al. 2005).
Fifth, the lack of brown dwarf companions to stars in wide binaries
equals the lack of free-floating brown dwarfs relative to stars
(Luhman et al. 2005). This implies that free floating brown dwarfs
form the same way as binary brown dwarfs.

\section{Mass Functions of Pre-Stellar Cores}

An important clue to the IMF is that the core mass function is often
similar to the stellar IMF, both in the slope at the high mass end
and in the turnover around $1\;M_\odot$. Table 1 lists observations
of core mass functions for low masses and similarly steep mass
functions for high mass cores. It also lists mass functions for high
mass cores that have a shallow slope, like the slope of the GMC mass
function.

\begin{deluxetable}{lllllll}
\tabletypesize{\scriptsize} \tablewidth{0pt}
\tablecaption{Observations of Proto-Stellar Core Mass Functions}
\tablehead{\colhead{Reference} &\colhead{Region} &\colhead{Method}
&\colhead{Number} &\colhead{Density} &\colhead{Mass Range}&\colhead{Slope}\\
&&&\colhead{of Cores} &\colhead{cm$^{-3}$}
&\colhead{$M_\odot$} &\colhead{$\Gamma$}}
\startdata
{\bf Low Mass Cores}&&&&&&\\
Motte et al. 1998    &$\rho$ Oph    &1.3mm       &58   &$10^6$     &$0.5-3$    &$-1.5$\\
Testi \& Sargent 1998&Serpens       &1.3mm       &26   &$10^7$     &$0.3-30$   &$-1.1$\\
Johnstone et al. 2000&$\rho$ Oph    &850mm       &55   &$10^{6.5}$ &$0.02-6.3$ &$-1$ to $-1.5$\\
Coppin et al. 2000   &Orion A North &450mm/850mm &67   &$10^5$     &$0.1-100$  &$-0.5$(shallow)\\
Kerton et al. 2001   &KR140         &450/850mm   &22   &$10^{4.3}$ &$0.5-130$  &$-0.5$
(shallow)\\
Johnstone et al. 2001&Orion B       &850mm       &75   &$10^5$     &$0.2-12.3$ &$-1.5$\\
Bontemps et al. 2001 &$\rho$ Oph    &6.7mm/14.3mm&123 Cl.II &...  &$0.02-3$   &$-1.7$\\
Motte et al. 2001    &NGC 2069/2071 &450mm/850mm &70   &$10^7$     &$0.3-5$    &$-1.1$\\
Sandell et al. 2001  &NGC 1333      &450mm/850mm &33   &$10^7$     &$0.03-1$   &$-0.4$ (near pk
only)\\
Tachihara et al. 2002&Tau, Oph, Lupus&&&&&\\
&L1333, CorAust,&&&&&\\
&Coalsack, Pipe neb.              &C$^{18}$O   &174  &$10^4$     &$1-400$  &$-1.5$ to $-2.6$\\
Onishi et al. 2002   &Taurus        &N$^{13}$CO$^+$&44 &$10^5$     &$3.5-20$   &$-1.5$\\
Tothill et al. 2002  &M8            &450/850/1.3mm &37&$10^{4.5}$ &$0.5-20$   &$-0.7$ (shallow)\\
Stanke et al. 2006 &Oph           &1.2mm       &111  &$10^5$     &$1-8$       &$-1.6$\\
Enoch et al. 2006  &Perseus       &1.1mm       &122  &$10^5$     &$1-30$      &$-1.6$\\
Johnstone et al. 2006&Orion B S   &450mm/850mm &57   &$10^6$     &$3-30$      &$-1.5$\\
Johnstone \& Bally 2006&Orion A S &450mm/850mm &71   &$10^6$     &$0.3-22$    &$-1$\\
Chi \& Park 06      &Polaris       &$^{13}$CO   &105  &$10^3$     &$0.1-10$    &$-0.91$ (low density)\\
Young et al. 2006  &Oph           &1.1mm       &44   &$10^6$     &$0.24-3.9$  &$-1.1$\\
Reid \& Wilson 2006&M17           &450mm/850mm &100  &$10^5$     &$0.8-120$   &$-0.5$ to $-0.9$\\
Kirk et al. 2006   &Perseus       &850mm/extinction&58 &$10^5$   &$0.3-5$     &$-2$\\
Alves et al. 2007  &Pipe Nebula   &extinction  &159  &$10^4$     &$0.5-28$    &$-1.4$\\
Ikeda et al. 2007  &Orion A       &H$^{13}$CO$^+$&236&$10^{4.3}$ &$0.5-2$     &$-1.5$\\
Nutter et al. 2007&Orion NS active&450mm/850mm &393  &$10^8$     &$0.1-40$    &$-1.2$\\
Li et al. 2007     &Orion S quiescent&sub-mm   &51   &$10^7$     &$0.1-46$    &$+0.15$\\
Walsh et al. 2007  &NGC 1333      &N$^2$H$^+$  &93   &$10^6$     &$0.05-2.5$  &$-1.4$\\
Massi et al. 2007  &Vela Cloud D  &1.2mm       &29   &$10^5$     &$0.04-88$   &$-0.45$ (shallow)\\
{\bf High Mass Cores}&&&&&&\\
Shirley et al. 2003&many sources  &CS          &57   &$10^5$     &$10^2-10^4$ &$0.91$\\
Reid \& Wilson 2005&NGC 7538      &450mm/850mm &67   &$10^5$     &$10^2-10^{3.5}$ &$-1$\\
Rathborne et al. 2006&IR Dark Clouds&1.2mm     &120  &$10^{4.5}$ &$10-10^{3.3}$&$-1.1$\\
{\bf Cluster Mass Slope}&&&&&&\\
Moore et al. 2007  &W3            &850mm       &316     &$10^5$  &$13-2500$    &$-0.8$\\
Beltran et al. 2006&Southern sources&IRAS      &235 &$10^6$ &$10^2-10^{3.6}$   &$-0.9$\\
&&&&&$10-120$&$-0.5$\\
Munoz et al. 2007 &NGC 6334       &1.2mm       &181  &           &$3-6000$     &$-0.6$
\enddata
\end{deluxetable}

There are several key properties of the pre-stellar cores that have
steep mass functions. First, they move at sub-virial, near sonic,
speeds, which implies that they do not accrete much in the
Bondi-Hoyle fashion. Measured velocity dispersions are $\sim0.17$ km
s$^{-1}$ (Di Francesco et al. 2004) or $\sim0.4$ km s$^{-1}$
(Andr\'e et al. 2007), and so on (Belloche et al. 2001; Walsh et al.
2004, 2007; Jorgensen et al. 2007; Kirk et al. 2007).

Second, they resemble Bonner-Ebert spheres, which means that they
are pressure-bound and self-gravitating, perhaps stable, although
sometimes there is evidence for inflow (e.g., Andr\'e et al. 2007).
Note that some theoretical derivations of the characteristic mass
are dynamic, which means they occur during the collapse and not in
equilibrium. For example, in Larson's (2005) model, the
characteristic mass in the IMF is the thermal Jeans mass at an
inflection point in the equation of state. This inflection is
important during the collapse as it determines the point where
subfragmentation stops (in the soft part of the equation of state)
and the existing fragments either keep their mass or grow by
accretion (in the hard part of the equation of state).

Third, the mass fractions of the pre-stellar cores inside their
clouds are $\sim2$ to 10\%, like the average star formation
efficiency. This implies the observed cores are evolving toward
individual stars or binary systems and nearly all the stars that
will ever form in the cloud are currently observed in the form of
cores.

Fourth, the cores cluster together hierarchically, like the young
stars they will eventually form (Johnstone et al. 2000, 2001; Enoch
et al. 2006; Young et al. 2006).

All of the pre-stellar cores that are observed with a particular
method in a region have about the same internal density, which is
the density where that method is most sensitive to finding them.
Overall, the pre-stellar cores have a  wide range of densities,
considering different methods and regions. The densities in the
table range from $10^4$ to $10^7$ cm$^{-3}$, even though the mass
functions are all about the same.

These observations suggest that the stellar IMF is determined in the
cloudy phase by gas processes, prior to significant collapse. This
differs from the IMF explanation in Bonnell et al. (2007), where the
stellar masses are determined by accretion onto existing cores whose
initial core mass function is not important and may even be a delta
function.  Bonnell et al. point out, however, that most pre-stellar
cores are not strongly self-gravitating, and they suggest that the
cores may not even form stars, or at least not on a one-to-one basis
with a preservation of the mass function.

Steep core mass functions are sometimes found for higher mass cores
too. The lower part of Table 1 lists the observations of these.
Because the cluster mass function has $\Gamma\sim1$, these cores
could form clusters.  Other mass functions at high mass are
shallower than this, like the GMC mass function. Examples of these
shallow functions are at the bottom of the table.

There are many detailed and interesting results about pre-stellar
core mass functions that have some bearing on the star formation
process. For example, the mass function for class II sources in
Ophiuchus is the same as the mass function for pre-stellar cores
there (Bontemps et al. 2001). This illustrates well the picture that
cores evolve into stars. The core definition most likely depends on
resolution, however. Enoch et al. (2007) found that the Serpens core
mass function is flatter (slope of $-1.6$) than the Perseus and
Ophiuchus mass functions (slopes of $-2.1$), all observed with the
same technique, and the distance to Serpens is the smallest. Yet all
three surveys have the same core size distribution relative to the
beam size. Thus, smaller cores are inferred for the closest regions.

Ikeda et al. (2007) studied Orion A with H$^{13}$CO$^+$ at a
characteristic density of $10^{4.3}$ cm$^{-3}$. They got core mass
functions with a turnover at low mass like the IMF, which is similar
to what other prestellar core studies found (as listed in Table 1),
but they got these turnovers only when they did not consider
blending corrections. They got a straight power law mass function
with no turnover when they included blending corrections. They
concluded from this that the turnover at low mass, which is one of
the key features that makes a pre-stellar core mass function look
like a stellar IMF, is the result of blending and source confusion.
If this is true in general, then we should not conclude without
further study that pre-stellar cores evolve into stars on a
one-to-one basis.

Another curious result is that the starless cores seen at 1.2mm in
Ophiuchus have mass segregation (Stanke et al. 2006): the mass
functions look the same in the inner and outer regions, but they
extend to higher mass in the inner regions.

Young et al. (2006) and Enoch et al. (2007) found that pre-stellar
cores are spatially correlated, which means they are hierarchically
clustered in the same way that stars are. This is further evidence
they will turn into stars without moving much further, considering
their velocity dispersions are low (see above).

Chi \& Park (2006) studied the polaris flare, which is a weakly
self-gravitating (diffuse) cloud visible in $^{12}$CO and $^{13}$CO
at a density of $\sim10^3$ cm$^{-3}$. The cores in this cloud have a
steep mass function too, $\Gamma=-0.91\pm0.13$. This suggests that
steep, Salpeter-like mass functions are made independently of the
star formation processes, long before self-gravity is important in
the gas. They seem to be characteristic of turbulence. For example,
a three-dimensional region with a Kolmogorov power-law power
spectrum and a log-normal density probability distribution function
has a mass function for clumps that is shallow at a low threshold
density and steep at a high threshold density (Elmegreen 2002;
Elmegreen et al. 2006).  The high-density mass function has the same
slope as the Salpeter IMF and the low-density mass function has the
same slope as the GMC mass function. Thus the mass functions for low
density clouds and pre-stellar cores may be sampled from the same
scale-free density distribution, but sampled at different density
thresholds.

\section{IMF Origins}

My best guess at the present time is that the IMF is a superposition
of several processes (Elmegreen 2004). There is an isolated
star-formation mode which is dominated by fragmentation near the
thermal Jeans mass, $M_J$. There is a dense cluster mode which adds
to this some extra accretion and clump coagulation to boost up the
high-mass component relative to that in the isolated mode. And
finally, there is the tiny dense-core mode where star mass is
heavily influenced by ejections from multiple systems and ablation
due to nearby stars. These latter processes modulate the isolated
mass function in the brown dwarf and sub-$M_J$ range.

Starbursts and major mergers (forming elliptical galaxies) could
have more of the dense cluster mode, which should give them slightly
flatter IMFs, as is sometimes observed. Low surface brightness
galaxies, dwarf galaxies with low ISM pressures, and quiescent
regions like the Taurus clouds have less of this dense cluster mode
and slightly steeper IMFs. This difference is not just a
temperature, pressure, or $M_J$ effect. It is the result of
different combinations of distinct physical processes.

\section{Maximum Stellar Mass}

\begin{figure}[h]
\centerline{
\includegraphics[width=250pt]{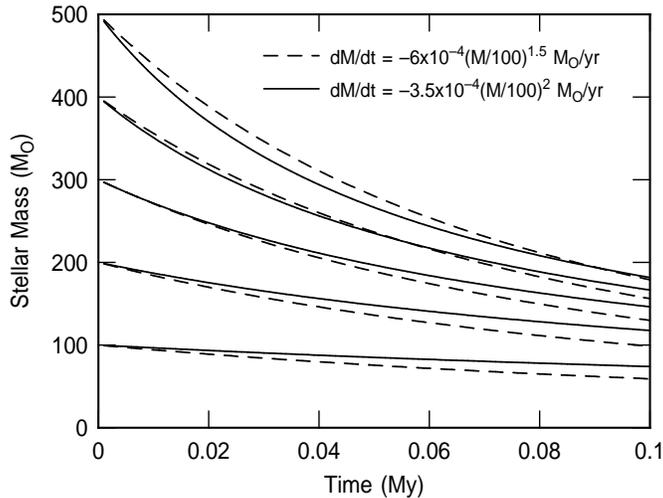}
} \caption{\label{imf4_sst} A model of stellar mass versus time when
a strong wind is present. The processes that determine the IMF may
be able to form $500\;M_\odot$ or larger ``protostars'', but these
objects will not show up as real stars if the mass loss rate is as
high as $\sim10^{-3} \;M_\odot$ yr$^{-1}$.}
\end{figure}

The IMF appears to have a maximum stellar mass of $120-150\;M_\odot$
(Weidner \& Kroupa 2004; Oey \& Clarke 2005; Koen 2006). This may
not be the maximum mass as far as the star formation process is
concerned, however. The IMF could in principle go to much higher
masses, but by the time the star appears as a pre-main sequence or
main-sequence object, it has lost so much mass via intense winds
that $\sim150\;M_\odot$ is all we can see.  Figure \ref{imf4_sst}
shows the time-dependent masses of stars that have the mass loss
rates indicated. Mass loss rates are assumed to depend on the
remaining mass. The initial masses are taken to be large, from
$100\;M_\odot$ to $500\;M_\odot$, but after only $10^5$ yrs, they
have all come down to the $100-200\;M_\odot$ mass range. This is the
time when they are likely to appear. The assumed mass loss rates are
not unreasonably high for young O-type stars.

\section{Summary}

The characteristic mass and slope of the IMF both show variations,
mostly in extreme environments. The characteristic mass increases
and/or the slope flattens a little ($\Gamma\sim1$ compared to the
Salpeter IMF which has $\Gamma=1.35$) in several regions that formed
at high redshift: the Milky Way and M31 bulges, the Milky Way
extremely metal-poor stars, elliptical galaxies, clusters of
galaxies, and other high redshift star formation. Perhaps the common
physical process is that these regions formed by major mergers and
starbursts in the early universe.  The Milky Way nucleus also
appears to have a somewhat flattened IMF, and a local bright rim of
presumably triggered star formation seems to have a high
characteristic mass. On the other hand, the IMFs for modern
starburst galaxies or super star clusters do not seem to be much
different from the Salpeter function. Perhaps they are not extreme
enough.

At the other extreme, dwarf galaxies and low surface brightness
galaxies have slightly steeper IMFs, $\Gamma\sim-1.6$. The field
regions between clusters and OB associations also have
systematically steeper IMFs.

Of the 3 models for the IMF discussed here (top down, bottom up,
interruption), the top-down model has the most direct evidence. This
evidence comes from the mass distribution functions of pre-stellar
cores and stars and from the slow random motions of pre-stellar
cores. Pre-stellar core mass functions resemble the IMF, which
implies that the IMF may be determined in the gas phase, with
possible modifications at high and low mass from stellar-scale
processes, such as fragmentation, collisions, and system ejections.


\begin{references}{}

\reference{} Alonso-Herrero, A., Engelbracht, C. W., Rieke, M. J.,
Rieke, G. H. \& Quillen, A. C. 2001, ApJ, 546, 952

\reference{} Alves, J., Lombardi, M., \& Lada, C. J. 2007, A\&A, 462, L17

\reference{} Andr\'e, Ph., Belloche, A., Motte, F., \& Peretto, N.
A\&A, 472, 519

\reference{} Ascenso, J., Alves, J., Beletsky, Y., \& Lago, M. T. V.
T. 2007, A\&A, 466, 137

\reference{} Ballero, S. K., Kroupa, P., \& Matteucci, F. 2007a,
A\&A, 467, 117

\reference{} Ballero, S. K., Matteucci, F., Origlia, L., \& Rich, R.
M. 2007b, A\&A, 467, 123

\reference{} Bate, M.R., Bonnell, I.A., \& Bromm, V. 2002, MNRAS,
332, L65

\reference{} Bate, M.R., \& Bonnell, I.A. 2005, MNRAS, 356, 1201

\reference{} Barrado y Navascu\'es, D., Stauffer, J.R., Bouvier, J.,
\& Mart\'in, E.L. 2001, ApJ, 546, 1006

\reference{} Bastian, N., \& Goodwin, S.P. 2006, MNRAS, 369, L9

\reference{} Belikov, A. N., Kharchenko, N. V., Piskunov, A. E. \&
Schilbach, E. 2000 A\&A, 358, 886

\reference{} Belloche, A., Andr\'e, P., \& Motte, F. 2001, in From
Darkness to Light: Origin and Evolution of Young Stellar Clusters,
ASP Conference Proceedings, Vol. 243. eds. T. Montmerle \& P.
Andr\'e, (San Francisco: ASP), p.313

\reference{} Beltr\'an, M. T., Brand, J., Cesaroni, R., Fontani, F., Pezzuto, S., Testi, L., \& Molinari, S. 2006, A\&A, 447, 221

\reference{} Bonnell, I.A., Larson, R.B., \& Zinnecker, H. 2007,
Protostars and Planets, ed. B. Reipurth, et al.\ (Tucson: Univ. of
Arizona), p. 149

\reference{} Bontemps, S. et al. 2001, A\&A, 372, 173

\reference{} Boss, A.P. 2001, ApJ, 551, L167

\reference{} Bouvier, J., Stauffer, J. R., Martin, E. L., Barrado y
Navascu\'es, D., Wallace, B., \& Bejar, V. J. S. 1998, A\&A, 336,
490

\reference{} Brice\~no, C., Luhman, K. L., Hartmann, L., Stauffer,
J. R. \& Kirkpatrick, J. D. 2002, ApJ, 580, 317

\reference{} Chandar, R., Leitherer, C., Tremonti, C.A., Calzetti,
D., Aloisi, A., Meurer, G.R., \& de Mello, D. 2005, ApJ, 628, 210

\reference{} Chi, S., \& Park, Y.-S. 2006, JKAS, 39, 9

\reference{} Chiosi, C. 2000, A\&A, 364, 423

\reference{} Clark, P.C. \& Bonnell, I.A. 2005, MNRAS, 361, 2

\reference{} Coppin, K. E. K., Greaves, J. S., Jenness, T., \&
Holland, W. S. 2000, A\&A, 356, 1031

\reference{} Dav\'e, R. 2008, MNRAS, on-line early edition, Feb,
2008

\reference{} de Marchi, G., Paresce, F., Straniero, O., \& Prada
Moroni, P.G. 2004, A\&A, 415, 971

\reference{} de Wit, W. J., Testi, L., Palla, F. \& Zinnecker, H.
2005, A\&A, 437, 247

\reference{} Di Francesco, J., Andr\'e, P., \& Myers, P.C. 2004,
ApJ, 617, 425

\reference{} Elmegreen, B.G. 2002, ApJ, 564, 773

\reference{} Elmegreen, B.G. 2004, MNRAS, 354, 367

\reference{} Elmegreen, B.G., \& Scalo, J. 2006, ApJ, 636, 149

\reference{} Elmegreen, B.G., Elmegreen, D.M., Chandar, R.,
Whitmore, B., \& Regan, M. 2006, ApJ, 644, 879

\reference{} Elmegreen, B.G., Klessen, R., \& Wilson, C. 2008, ApJ,
submitted

\reference{} Enoch, M.L. et al. 2006, ApJ, 638, 293

\reference{} Fardal, M.A. Katz, N., Weinberg, D.H., \& Dav\'e, R.
2007, MNRAS, 379, 985

\reference{} Getman, K.V., Feigelson, E.D., Garmire, G., Broos, P.,
\& Wang, J. 2007, ApJ, 654, 316

\reference{} Gonz\'alez Delgado, R. M., P\'erez, E. 2000 MNRAS, 317,
64

\reference{} Goodwin, S.P., \& Whitworth, A. 2007, A\&A 466, 943

\reference{} Gouliermis, D., Brandner, W., \& Henning, Th. 2005,
ApJ, 623, 846

\reference{} Harayama, Y., Eisenhauer, F., \& Martins, F. 2007,
astro-ph/0710.2882

\reference{} Ho, L. C. \& Filippenko, A. V. 1996 ApJ, 466, L83

\reference{} Hodge, P. 1986, PASP, 98, 1113

\reference{} Holtzman, J. A., Watson, A. M., Baum, W.A., Grillmair,
C.J., Groth, E.J., Light, R.M., Lynds, R., O'Neil, E.J., Jr. 1998,
AJ, 115, 1946

\reference{} Hoversten, E.A., Glazebrook, K. 2007,
astro-ph/0711.1309

\reference{} Ikeda, N., Sunada, K., \& Kitamura, Y. 2007, ApJ, 665,
1194

\reference{} Jappsen, A.-K., Klessen, R.S., Larson, R.B., Li, Y., \&
MacLow M.-M. 2005, A\&A, 435, 611

\reference{} Johnstone, D., Wilson, C.D., Moriarty-Schieven, G.,
Joncas, G., Smith, G., Gregersen, E., \& Fich, M. 2000, 545, 327

\reference{} Johnstone, D., Fich, M., Mitchell, G.F., \&
Moriarty-Schieven, G. 2001, ApJ, 559, 307

\reference{} Johnstone, D., Matthews, H., \& Mitchell, G.F. 2006, ApJ, 639, 259

\reference{} Johnstone, D., \& Bally, J. 2006, ApJ, 653, 383

\reference{} J{\o}rgensen, J.K., Johnstone, D., Kirk, H., \& Myers,
P.C. 2007, ApJ, 656, 293

\reference{} Kerton, C. R., Martin, P. G., Johnstone, D., \& Ballantyne, D. R. 2001, ApJ, 552, 610

\reference{} Kim, S.S., Figer, D.F., Kudritzki, R.P., \& Najarro, F.
2006, ApJ, 653, L113

\reference{}  Kirk, H., Johnstone, D., \& Di Francesco, J. 2006, ApJ, 646, 1009

\reference{} Kirk, H., Johnstone, D., \& Tafalla, M. 2007, ApJ, 668,
1042

\reference{} Koen, C. 2006, MNRAS, 365, 590

\reference{} Komiya, Y., Suda, T., Habe, A., \& Fujimoto, M.Y. 2007,
astro-ph/0710.4374

\reference{} Kroupa, P., Tout, C.A., Gilmore, G. 1993, MNRAS, 262,
545

\reference{} Kroupa, P., \& Bouvier, J. 2003, MNRAS, 346, 369

\reference{} Larsen, S.S., Brodie, J.P., Elmegreen, B.G., Efremov,
Y.N., Hodge, P.W. \& Richtler, T.  2001, ApJ, 556, L801

\reference{} Larson, R.B. 2005, MNRAS, 359, 211

\reference{} Lee, H.-C., Gibson, B.K., Flynn, C., Kawata, D., \&
Beasley, M.A. 2004, MNRAS, 353, 113

\reference{} Li, D., Velusamy, T., Goldsmith, P. F., \& Langer, W.D. 2007, ApJ, 655, 351

\reference{} Li, P.S., Norman, M.L., Mac Low, M.-M., \& Heitsch, F.
2004, ApJ, 605, 800

\reference{} Loewenstein, M., \& Mushotsky, R.F. 1996, ApJ, 466, 695

\reference{} Lucke, P. B., \& Hodge, P. W. 1970, AJ, 75, 171

\reference{} Luhman, K.L. 2004a, ApJ, 617, 1216

\reference{} Luhman, K.L. 2004b, ApJ, 614, 398

\reference{} Luhman, K. L., McLeod, K. K., \& Goldenson, N. 2005,
ApJ ,623, 1141

\reference{} Luhman, K.L. 2006, ApJ, 645, 676

\reference{} Luhman, K.L. 2007, ApJS, 173, 104

\reference{} Luhman, K.L. \& Rieke, G.H., 1999, ApJ, 525, 440

\reference{} Luhman, K.L., Stauffer, J.R., Muench, A.A., Rieke,
G.H., Lada, E.A., Bouvier, J., \& Lada, C.J. 2003, ApJ, 593, 1093

\reference{} Luhman, K. L., Whitney, B. A., Meade, M. R., Babler, B.
L., Indebetouw, R., Bracker, S., \& Churchwell, E. B. 2006, ApJ,
647, 1180

\reference{} MacLow, M.-M., \& Klessen, R.S. 2004, Rv.Mod.Phys., 76,
125

\reference{} Maness, H., Martins, F., Trippe, S., Genzel, R.,
Graham, J. R., Sheehy, C., Salaris, M., Gillessen, S., Alexander,
T., Paumard, T., Ott, T., Abuter, R., \& Eisenhauer, F. 2007, ApJ,
669, 1024

\reference{} Martel, H., Evans, N.J., II \& Shapiro, P.R. 2006,
ApJS, 163, 122

\reference{} Mas-Hesse, J. M., \& Kunth, D. 1999, A\&A, 349, 765

\reference{} Massey, P. \& Hunter, D.A. 1998, ApJ, 493, 180

\reference{} Massey, P. 2002, ApJS, 141, 81

\reference{} Massey, P., Johnson, K.E., \& Degioia-Eastwood, K.
1995, ApJ, 454, 151

\reference{} Massi, F., de Luca, M., Elia, D., Giannini, T., Lorenzetti, D., \& Nisini, B. 2007, A\&A, 466, 1013

\reference{} McCrady, N., Gilbert, A. \& Graham, J.R. 2003, ApJ,
596, 240

\reference{} Mengel, S., Lehnert, M. D., Thatte, N. \& Genzel, R.
2002, A\&A, 383, 137

\reference{} Moore, T. J. T., Bretherton, D. E., Fujiyoshi, T.,
Ridge, N. A., Allsopp, J., Hoare, M. G., Lumsden, S. L., \& Richer, J. S. 2007, MNRAS, 379, 663

\reference{} Moraux, E., Bouvier, J., Stauffer, J. R., Barrado y
Navascu\'es, D., \& Cuillandre, J.-C. 2007, A\&A, 471, 499

\reference{} Moretti, A., Portinari, L., \& Chiosi, C. 2003, A\&A,
408, 431

\reference{} Motte, F., Andr\'e, P., \& Neri, R. 1998, A\&A, 336, 150

\reference{} Motte, F., Andr\'e, P., Ward-Thompson, D., \& Bontemps, S. 2001, A\&A, 372, L421

\reference{} Muench, A.A., Lada, E.A., Lada, C.J., \& Alves, J.
2002, ApJ, 573, 366

\reference{} Mu\~noz, D.J., Mardones, D., Garay, G., Rebolledo, D., Brooks, K., \& Bontemps, S. 2007, ApJ, 668, 906

\reference{} Muzerolle, J., Luhman, K.L., Brice\~no, C. Hartmann,
L., \& Calvet, N. 2005, ApJ, 625, 906

\reference{} Nagashima, M., Lacey, C. G., Baugh, C.M., Frenk, C.S.,
\& Cole, S. 2005a, MNRAS, 363, 1247

\reference{} Nagashima, M., Lacey, C. G., Okamoto, T., Baugh, C.M.,
Frenk, C.S., \& Cole, S. 2005b, MNRAS, 363, L31

\reference{} Nakamura, F. \& Li, Z.-Y. 2005, ApJ, 631, 411

\reference{} Nakamura, F., \& Li, Z.-Y. 2007, ApJ, 662, 395

\reference{} Nayakshin, S., \& Sunyaev, R. 2005, MNRAS, 364, L23

\reference{} Nutter, D., \& Ward-Thompson, D. 2007, MNRAS, 374, 1413

\reference{} Oey, M. S., \& Clarke, C. J 2005, ApJ, 620, L43

\reference{} Okumura, S., Mori, A., Nishihara, E., Watanabe, E. \&
Yamashita, T. 2000, ApJ, 543, 799

\reference{} Onishi, T., Mizuno, A., Kawamura, A., Tachihara, K., \& Fukui, Y. 2002, ApJ, 575, 950

\reference{} Padoan, P, Nordlund, A., Kritsuk, A.G., Norman, M.L.,
\& Li, P.S. 2007, ApJ, 661, 972

\reference{} Paresce, F. \& de Marchi, G. 2000, ApJ, 534, 870

\reference{} Parker, J.W., Hill, J.K., Cornett, R.H., Hollis, J.,
Zamkoff, E., Bohlin, R. C., O'Connell, R.W., Neff, S.G., Roberts,
M.S., Smith, A.M. \& Stecher, T.P. 1998 AJ, 116, 180

\reference{} Pasquali, A., de Marchi, G., Pulone, L., \& Brigas,
M.S. 2004, A\&A, 428, 469

\reference{} Paumard, T., Genzel, R., Martins, F., Nayakshin, S.,
Beloborodov, A. M., Levin, Y., Trippe, S., Eisenhauer, F., Ott, T.,
Gillessen, S., Abuter, R., Cuadra, J., Alexander, T., \& Sternberg,
A. 2006, ApJ, 643, 1011

\reference{} Pipino, A., \& Matteucci, F. 2004, MNRAS, 347, 968

\reference{} Portinari, L., Moretti, A., Chiosi, C., \&
Sommer-Larsen, J. 2004a, ApJ, 604, 579

\reference{} Preibisch, T., Brown, A.G.A., Bridges, T., Guenther, E.
\& Zinnecker, H. 2002, AJ, 124, 404

\reference{} Rana, N.C. 1987, A\&A, 184, 104

\reference{} Rathborne, J.M., Jackson, J.M., \& Simon, R. 2006, ApJ, 641, 389

\reference{} Reid, M.A., \& Wilson, C.D. 2005, ApJ, 625, 891

\reference{} Reid, M.A., \& Wilson, C.D. 2006, ApJ, 644, 990

\reference{} Reipurth, B., \& Clarke, C. 2001, AJ, 122, 432

\reference{} Renzini, A., Ciotti, L., D'Ercole, A., \& Pellegrini,
S. 1993, ApJ, 419, 52

\reference{} Romeo, A. D., Sommer-Larsen, J., Portinari, L., \&
Antonuccio-Delogu, V. 2006, MNRAS, 371, 548

\reference{} Sandell, G., Knee, L.B.G. 2001, ApJ, 546, L49

\reference{} Sanner, J., Altmann, M., Brunzendorf, J. \& Geffert, M.
2000, A\&A, 357, 471

\reference{} Scalo, J.M. 1986, Fund.Cos.Phys, 11, 1

\reference{} Scalo, J.M. 1998, in The Stellar Initial Mass Function,
ed. G. Gilmore, I. Parry, \& S. Ryan, Cambridge: Cambridge
University Press, p. 201

\reference{} Scholz, A., \& Jayawardhana, R. 2008, ApJ, 672, L49

\reference{} Selman, F. \& Melnick, J. 2005, A\&A, 443, 851

\reference{} Shirley, Y.L., Evans, N.J., II, Young, K.E., Knez, C., \& Jaffe, D.T. 2003, ApJS, 149, 375

\reference{} Slesnick, C.L., Hillenbrand, L.A. \& Massey, P. 2002,
ApJ, 576, 880

\reference{} Smith, L.J., Gallagher, J.S. 2001, MNRAS, 326, 1027

\reference{} Stanke, T., Smith, M. D., Gredel, R., \& Khanzadyan, T.
2006, A\&A, 447, 609

\reference{} Sternberg, A. 1998, ApJ, 506, 721

\reference{} Stolte, A., Brandner, W., Brandt, B., \& Zinnecker, H.
2006, AJ, 132, 253

\reference{} Tachihara, K., Onishi, T., Mizuno, A., \& Fukui, Y. 2002, A\&A 385, 909

\reference{} Testi, L., \& Sargent, A.I. 1998, ApJ, 508, L91

\reference{} Tilley, D. A., \& Pudritz, R.E. 2007, MNRAS, 382, 73

\reference{} Tornatore, L., Borgani, S., Matteucci, F., Recchi, S.,
\& Tozzi, P. 2004, MNRAS, 349, L19

\reference{} Tothill, N. F. H., White, G.J., Matthews, H. E., McCutcheon, W. H., McCaughrean, M. J., \& Kenworthy, M. A. 2002, ApJ, 580, 285

\reference{} \'Ubeda, L., Ma\'iz-Apell\'aniz, J., \& MacKenty, J.W.
2007a, AJ, 133, 917

\reference{} \'Ubeda, L., Ma\'iz-Apell\'aniz, J., \& MacKenty, J.W.
2007b, AJ, 133, 932

\reference{}  Umbreit, S., Burkert, A., Henning, T., Mikkola, S., \&
Spurzem, R. 2005, ApJ, 623, 940

\reference{} van Dokkum, P.G. 2008, ApJ, 674, 29

\reference{} Walsh, A.J., Myers, P.C., \& Burton, M.G. 2004, ApJ,
614, 194

\reference{} Walsh, A.J., Myers, P.C., Di Francesco, J., Mohanty, S.
Bourke, T.L., Gutermuth, R., \& Wilner, D. 2007, ApJ, 655, 958

\reference{} Wang, J., Townsley, L.K., Feigelson, E.D., Broos, P.S.,
Getman, K.V., Roman-Zuniga, C., \& Lada, E. 2007, astro-ph/0711.2024

\reference{} Weidner, C., \& Kroupa, P. 2004, MNRAS, 348, 187

\reference{} Whitworth, A.P., Boffin, H.M.J., \& Francis, N. 1998,
MNRAS, 299, 554

\reference{} Whitworth, A. P., \& Zinnecker, H. 2004, A\&A, 427, 299

\reference{} Whitworth, A.P., \& Goodwin, S.P. 2005, Memorie della
Societa Astronomica Italiana, 76, p.211


\reference{} Yasui, C., Kobayashi, N., Tokunaga, A.T., Saito, M.,
Tokoku, C. 2007, astro-ph/0801.0204

\reference{} Young, K.E. et al. 2006, ApJ, 644, 326

\reference{} Zackrisson, E., Bergvall, N., Marquart, T., Mattsson,
L., \&  \"Ostlin, G. 2005, in Starbursts: From 30 Doradus to Lyman
Break Galaxies, eds. R. de Grijs and R.M. Gonz\'alez Delgado, Ap\&
Sp.Sci.Lib., 329, p.86

\reference{} Zoccali, M., Cassisi, S., Frogel, J.A., Gould, A.,
Ortolani, S., Renzini, A., Rich, R. M., \& Stephens, A.W. 2000, ApJ,
530, 418


\end{references}
\end{document}